\documentclass[12pt]{iopart}
\usepackage{graphicx}
\usepackage{dcolumn}
\usepackage{bm}

\begin{document}

\title[TASEP with Junction]{Theoretical Investigation of Totally Asymmetric Exclusion Processes on Lattices with Junctions}
\author{Ekaterina Pronina\dag  and Anatoly B. Kolomeisky\dag\ddag}
\address{\dag Department of Chemistry, Rice University, Houston, TX 77005, USA}
\address{\ddag Department of Chemical and Biomolecular Engineering, Rice University, Houston, TX 77005, USA}

\begin{abstract}
Totally asymmetric simple exclusion processes on lattices  with junctions, where particles interact with hard-core exclusion and move on parallel lattice branches that at the junction combine  into a single lattice segment, are investigated. A simple approximate theory, that treats the correlations around the junction position in a mean-field fashion, is developed in order to calculate  stationary  particle currents, density profiles and a phase diagram. It is shown that there are three possible  stationary phases depending on the state of each of the lattice branch. At first-order phase boundaries, where the density correlations are important, a modified phenomenological domain-wall theory, that accounts for correlations, is introduced. Extensive Monte Carlo computer simulations are performed to investigate the system, and it is found that they are in excellent  agreement with theoretical predictions.  
\end{abstract}


\ead{tolya@rice.edu}

\maketitle

\section{Introduction}

Asymmetric simple exclusion processes (ASEP) have been  introduced  originally in 1968 as a theoretical model for a description of  kinetics of biopolymerization \cite{macdonald}. Although in last years  the area of application of  ASEP has been significantly broadened \cite{derrida93,derrida98,schutz}, and it now includes a road traffic flow analysis \cite{nagel96,lee}, polymer dynamics in dense medium \cite{schutz99}, and many other problems,  the major application of ASEPs remains  the modeling of various biophysical transport phenomena. In particular, simple exclusion processes have been used successfully to describe  protein synthesis \cite{shaw03,shaw04}, mRNA translation phenomena \cite{chou}, gel electrophoresis \cite{widom91}, a  motion of motor proteins along the cytoskeletal filaments \cite{klumpp03} and  the depolymerization of microtubules by  special enzymes \cite{klein05}.

ASEP are  discrete non-equilibrium models that describe  stochastic dynamics of multi-particle transport along  one-dimensional lattices. The lattices are finite and generally consist of $L \gg  1$ sites. Each site can be  occupied by a single particle or empty, and the particles interact only through the hard-core exclusion potential.  The  dynamics of ASEP is asymmetric, i.e., the particles can hop to the left or to the right but with different probabilities. In the simplest totally asymmetric simple exclusion process (TASEP) the particles move only in one direction. In this case the rules for the motion are the following. The particle at site $1 \le i < L$ can move one step forward if the site $i+1$ is empty. The particle can enter the lattice with the rate $\alpha$  if the first site is available, and it can leave the system  from the last site $i=L$ with the rate $\beta$. In the stationary-state limit of TASEP, the system  can be found in one of three phases depending on entrance, exit or bulk processes dominate the overall dynamics.

The unusual dynamic properties and phase behavior of ASEP and a wide range of applications in chemistry, physics and biology stimulated many theoretical studies of asymmetric exclusion processes \cite{derrida93,derrida98,schutz}. There are several exact results for the steady-state properties  of ASEP for different update rules \cite{derrida98,schutz}, although most theoretical investigations utilize the approximate methods along with Monte Carlo computer simulations \cite{shaw03,shaw04,mirin03,klumpp04}. The coupling of several  exclusion processes has  been considered in the study of parallel-chain ASEP \cite{ps03,popkov04,pronina04} The combining of non-equilibrium exclusion processes with equilibrium particle association/dissociation phenomena led  to unusual phenomena of  localizations of density shocks  \cite{parm03,parm04,popkov03,evans03,levine04}. However, despite the differences in the specific dynamic rules and stationary properties of specific ASEP, the microscopic origin of unusual dynamic properties and phase behavior can be well understood with the help of a phenomenological domain wall theory \cite{kolomeisky98,schutz}.

The majority of the investigated asymmetric exclusion processes deal with the particle movement along the one-channel lattices. Although the one-channel approach describes many situations in the biophysical processes, the more realistic description of the cellular transport requires an extension of original ASEP to include the possibility of a transport on the lattices with a more complex geometry. For example, consider the motor proteins kinesins that move vesicles and organelles along the microtubules and play important role in cellular transport \cite{howard,klumpp03}. Microtubules are made of parallel linear polymers, called protofilaments, that are arranged circumferentially. It is known that kinesins walk only on single protofilaments. However, experiments \cite{chretien} indicate  that the number of protofilaments may vary, at least for {\it in vitro} conditions, and it indicates  the existence of junctions and other lattice defects. Such defects might lead to  motor proteins crowding phenomena that are responsible for many human diseases \cite{aridor01}. These observations  suggest an importance of investigation of  the asymmetric exclusion processes on the lattices with junctions as a model for these complex biological transport phenomena. 

Recently  Brankov  et al. \cite{brankov04} have investigated TASEP on chains with a double-chain section in the middle by using an approximate theory and computer simulations. This corresponds to having two consecutive junctions on the lattice. Several stationary phases, the existence of which has not been expected, are found. In addition, the density profiles at phase boundaries and strong correlations between different lattice branches have been observed, but they could not be explained because the theory neglected the correlations. In this paper, we present a theoretical investigation of TASEP on the lattice with one junction. We develop a simple approximate theory that allows to calculate particle currents, density profiles and a  phase diagram at large times. At first-order phase boundaries, we  develop an extension of the  domain wall theory \cite{kolomeisky98} that accounts for correlations. This approach is then applied for explicit  calculations of density profiles. The theoretical predictions are compared with extensive Monte Carlo computer simulations.

The paper is organized as follows. In section 2 the model is introduced and theoretical calculations of stationary properties  are presented. In section 3 theoretical predictions are compared and discussed with  the results of Monte Carlo computer simulations. The final summary and conclusions are given in section 4.

\section{Theoretical Description} 

\subsection{Model}

We consider identical particles that move along the lattice with the junction positioned in the middle of the system as shown in Fig. 1. The system is out of equilibrium, and it has three equal-size branches, each containing $L$ sites. The particles can enter chain I or chain II with the rate  $\alpha$ if the first site at the corresponding  branch is available. Two chains merge together at site $L+1$ and form chain III. The particle can leave the system with the rate $\beta$: see Fig. 1. Inside of the lattice chains the particle can only move one step forward if the neighboring site is empty.  

\begin{figure}[ht] 
\centering
\includegraphics[scale=0.6, clip=true]{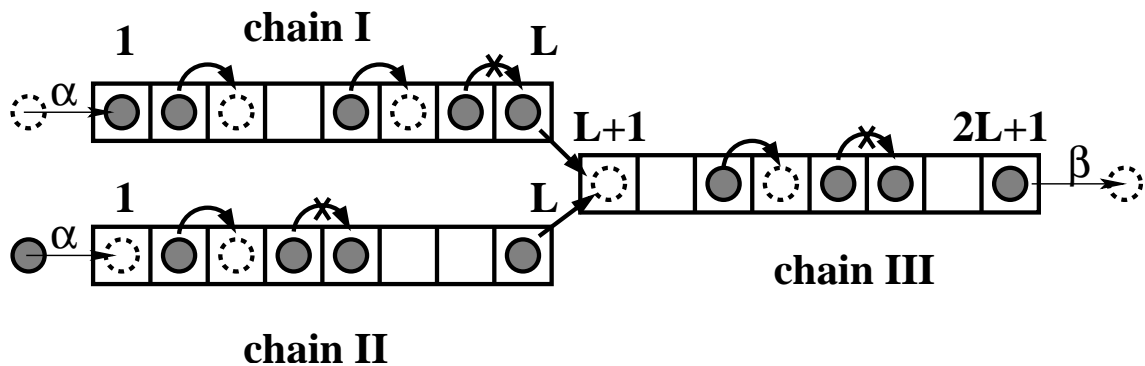}
\caption{\small{Schematic picture for the model for TASEP on the lattice with  a junction. Particles can enter  chain I or chain II with the  rate $\alpha$. At the site $L+1$ two lattice branches coalesce into chain III, from which particles can exit with the rate $\beta$. Arrows indicate the allowed transitions, while  crossed arrows correspond to the prohibited moves.}}
\end{figure}

Without chain I or chain II, the junction disappears and the system reduces to a totally asymmetric simple exclusion process on the one-channel lattice, for which the full description of stationary properties, such as phase diagram, particle currents and density profiles, are known \cite{derrida93,derrida98,schutz}. This simplest model of TASEP on the lattice without junctions has three stationary phases. When the entrance into the system is a rate-limiting process, for $\alpha<1/2$ and $\alpha<\beta$, the system is found in a low-density (LD) phase with the current and bulk density given by
\begin{equation}
J_{LD}=\alpha (1-\alpha), \quad \rho_{bulk,LD}= \alpha.
\end{equation}
If exit controls the dynamics of the system, for $\beta<1/2$ and $\beta<\alpha$, the stable stationary state for the system is a high-density (HD) phase with the following current and bulk density
\begin{equation}
J_{HD}=\beta (1-\beta), \quad \rho_{bulk,HD}= 1-\beta.
\end{equation}
Finally, for large entrance and exit rates ($\alpha>1/2$ and $\beta>1/2$), when the dynamics  is determined by bulk processes, the system is in a maximal-current (MC) phase with  
\begin{equation}
J_{MC}=1/4, \quad \rho_{bulk,MC}=1/2.
\end{equation}
Full density profiles for  systems of any size can be calculated explicitly \cite{derrida93}.

The microscopic origin of complex phase behavior and  unusual dynamic properties  of asymmetric exclusion processes can be explained via a phenomenological domain-wall (DW) theory \cite{kolomeisky98}. According to this approach, the domain wall is the boundary region between two possible stationary phases, and it moves through the system as a random walker with a speed determined by the currents and densities in two phases:
\begin{equation} \label{dw_vel}
v_{DW}=u_{+}-u_{-}=\frac{J_{+}-J_{-}}{\rho_{+}-\rho_{-}},
\end{equation}
where ``+'' (``$-$'') corresponds to the phase to the right (left) of the domain wall, and $u_{+}$ and $u_{-}$ give the domain wall rates  to hop to the right or left. For $v_{DW} >0$ ($u_{+} > u_{-}$) the domain wall moves to the right and the ``negative'' phase becomes a stationary state of the system, while for $v_{DW} <0$  ($u_{+} < u_{-}$) the domain wall  travels to the left and the ``positive'' phase wins over. On the phase boundaries the domain wall has equal probability to go forward or backward, i.e., $u_{+}=u_{-}$ and $v_{DW}=0$. As a result,  the density profiles are  linear. This is due to the fact that the domain wall can be found with equal probability at any position in the system.

\subsection{Theoretical calculations for stationary phases}

The overall  state of the system is specified by the nature of phases that might exist in each of the lattice branches. Since  only three  stationary phases can be found  in each lattice chain (HD, LD or MC), the total number of possible stationary phases in the system with the junction is equal to $3^{3}=27$. However, because of a symmetry, chains I and II should have identical phases  (see Fig. 1), and the overall state of the system is determined by phases in the chain I (or II) and the chain III.  Then the number of possible stationary states reduces to only $3^{2}=9$. 

The overall stationary current passing through the system can be written as
\begin{equation} \label{stat_current}
J_{overall}=J_{III}=J_{I}+J_{II}=2 J_{I}.
\end{equation} 
It suggests that chain I and II cannot have the the maximal-current phase with $J=1/4$,  because the maximal possible current through the system is just equal to 1/4. Thus, there are only 6 possible stationary phases: (LD,LD), (LD,HD), (LD,MC), (HD,LD), (HD,HD) and (HD,MC), where in the expression (A,B) A describes the phase in the chains I and II, while B corresponds to the phase in the chain III.

The junction  introduces an inhomogeneity in the system and it makes impossible to solve the large-time dynamics exactly \cite{mirin03,kol98}. However, TASEP on the lattice with the junction can be mapped into 3 coupled homogeneous asymmetric exclusion processes as shown in Fig. 2, for which the approximate description can be developed \cite{mirin03,kol98}. In order to obtain the dynamic properties of the system explicitly, in the simplest approximation,  we assume that there are no correlations in the occupation of the sites before and after the junction, i.e.,
\begin{equation}
J_{overall}=2 J_{junction}= 2 < \tau_{L}(1-\tau_{L+1}) > \approx 2 < \tau_{L} > (1-< \tau_{L+1} >),
\end{equation} 
where $< \tau_{L}> =\rho_{L} $ is the  probability to occupy the site $L$ on the chain I or II, and $< \tau_{L+1}> =\rho_{L+1} $ is the average particle  density at the site $L+1$ of the chain III. The effective rates $\alpha_{eff}$ and $\beta_{eff}$ (see Fig. 2) can be expressed in terms of the particle densities at the sites near the junction, 
\begin{equation} \label{eff_rates_def}
\alpha_{eff}=2\rho_{L}, \quad  \beta_{eff}=1-\rho_{L+1}.
\end{equation}

\begin{figure}[h] 
\centering
\includegraphics[scale=0.6, clip=true]{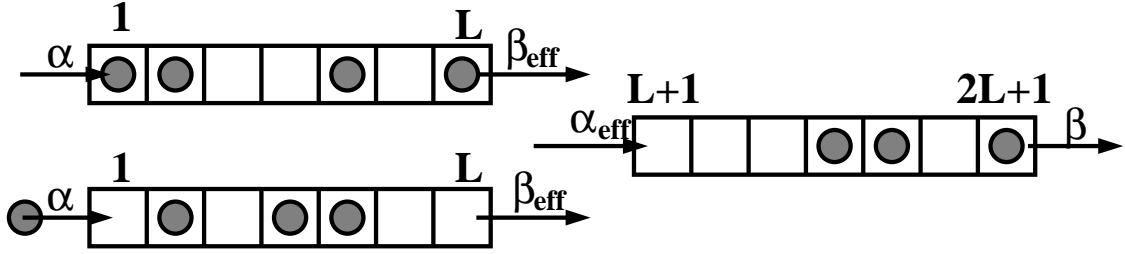}
\caption{\small{The totally asymmetric simple exclusion process on the lattice with the junction can be mapped into three  homogeneous TASEP coupled at the ends. The effective rate $\beta_{eff}$ describes the process of exiting from the chains I or II, while the effective rate $\alpha_{eff}$ corresponds to the entrance into the chain III. }}
\end{figure}

Now we can investigate the existence of different stationary phases. Consider first (LD,LD) phase, which can be specified by  the following conditions,
\begin{equation} \label{LD,LD}
 \alpha<1/2, \quad \alpha < \beta_{eff}; \quad \alpha_{eff}<1/2, \quad  \alpha_{eff} < \beta.
\end{equation}
The stationary currents and bulk densities are given by
\begin{equation}
J_{I}=\alpha (1-\alpha), \quad  \rho_{I,bulk}=\alpha , \quad J_{III}=\alpha_{eff} (1-\alpha_{eff}), \quad  \rho_{III,bulk}=\alpha_{eff}.
\end{equation}
Using the expression (\ref{stat_current}) for currents, the effective rate  $\alpha_{eff}$ can be expressed in terms of the  entrance rate $\alpha$,
\begin{equation}\label{alpha_eff_LDLD}
\alpha_{eff} = \frac{1-\sqrt{1-8\alpha(1-\alpha)}}{2}. 
\end{equation}
Thus $\alpha_{eff}$ always satisfy the conditions (\ref{LD,LD}), however, this equation yields  physically reasonable values of the effective entrance rate only when the term in the square root is positive, i.e., 
\begin{equation}\label{alpha_LDLD}
\alpha  < \frac{1}{2}-\frac{\sqrt{2}}{4} \approx 0.146.
\end{equation}
Because $\rho_{L+1}=\alpha_{eff}$ in (LD,LD) phase, the effective exit rate, as follows from Eq. (\ref{eff_rates_def}), is given by $\beta_{eff}=1-\alpha_{eff}$. Then the condition $\alpha < \beta_{eff}$ is true for all values of parameters. Thus the system is in (LD,LD) phase  when
\begin{equation} \label{beta_LDLD}
\beta > \frac{1-\sqrt{1-8\alpha(1-\alpha)}}{2} \quad \mbox{and  } \alpha  < \frac{1}{2}-\frac{\sqrt{2}}{4}.
\end{equation}

For the (LD,HD) phase the conditions of existence can be written as:
\begin{equation} \label{LD,HD}
 \alpha<1/2, \quad \alpha < \beta_{eff}; \quad \beta < 1/2, \quad \beta < \alpha_{eff};
\end{equation}
while the stationary currents and bulk densities are
\begin{equation}
J_{I}=\alpha (1-\alpha), \quad  \rho_{I,bulk}=\alpha , \quad J_{III}=\beta (1-\beta), \quad  \rho_{III,bulk}=1-\beta. 
\end{equation}
Because the particle current is stationary, Eq. (\ref{stat_current}) implies that
\begin{equation} \label{beta_LDHD}
\beta = \frac{1-\sqrt{1-8\alpha(1-\alpha)}}{2}, \quad \mbox{and  } \alpha  < \frac{1}{2}-\frac{\sqrt{2}}{4}.
\end{equation}
These expressions  describe the parameter's space for (LD,HD) phase.

Similar analysis can be performed for (LD,MC) phase. The allowed parameters for this phase are specified by
\begin{equation} \label{LD,MC}
 \alpha<1/2, \quad \alpha < \beta_{eff};  \quad  \alpha_{eff} >1/2, \quad \beta > 1/2.
\end{equation}
The currents and bulk densities have the following values: 
\begin{equation}
J_{I}=\alpha (1-\alpha), \quad  \rho_{I,bulk}=\alpha , \quad J_{III}=1/4, \quad  \rho_{III,bulk}=1/2. 
\end{equation}
Then the currents in the chain I and II are $J_{I}=J_{II}=1/8$, which implies that
\begin{equation} 
\alpha = \frac{1}{2}-\frac{\sqrt{2}}{4}. 
\end{equation}
This equation along with the condition $\beta > 1/2$ fully determines the region of existence of (LD,MC) phase.

The situation is very different for (HD,LD) phase for which
\begin{equation} \label{HD,LD}
 \beta_{eff} < 1/2, \quad \alpha > \beta_{eff}; \quad \alpha_{eff} < 1/2, \quad  \alpha_{eff} < \beta.
\end{equation}
The corresponding equations for the currents and bulk densities are
\begin{equation}
J_{I}=\beta_{eff} (1-\beta_{eff}), \quad  \rho_{I,bulk}=1-\beta_{eff} , \quad J_{III}=\alpha_{eff} (1-\alpha_{eff}), \quad  \rho_{III,bulk}=\alpha_{eff}.
\end{equation}
It is known that in this phase $\rho_{L}=1-\beta_{eff}$ and $\rho_{L+1}=\alpha_{eff}$. Then from Eq. (\ref{eff_rates_def}) it can be shown that
\begin{equation} 
\alpha_{eff}=2(1-\beta_{eff}), \quad \mbox{and } \beta_{eff}=1-\alpha_{eff}.
\end{equation}
However, these two equations have no real solutions together, and therefore (HD,LD) phase cannot exist for any value of the entrance rate $\alpha$ and the exit rate $\beta$.  

The (HD,HD) phase is determined from the conditions
\begin{equation} \label{HD,HD}
 \beta_{eff} < 1/2, \quad \beta_{eff} < \alpha ; \quad \beta < 1/2, \quad \beta < \alpha_{eff};
\end{equation}
The stationary properties of this phase are given by
\begin{equation}
J_{I}=\beta_{eff} (1-\beta_{eff}), \quad  \rho_{I,bulk}=1-\beta_{eff} , \quad J_{III}=\beta (1-\beta), \quad  \rho_{III,bulk}=1-\beta. 
\end{equation}
The stationary condition for the particle currents [see Eq.(\ref{stat_current})] helps to determine the effective exit rate constant,
\begin{equation} 
\beta_{eff}=\frac{1-\sqrt{1-2\beta(1-\beta)}}{2}.
\end{equation}  
After combining this result with  the set of  phase existence requirements (\ref{HD,HD}) we obtain the final conditions for (HD,HD) phase:
\begin{equation}
\beta<\frac{1}{2}, \quad  \mbox {if } \alpha > \frac{1-\sqrt{1-2\beta(1-\beta)}}{2};
\end{equation}  
and
\begin{equation}
\beta < 1/2, \quad \mbox{ if } \alpha<1/2.
\end{equation}

The last possible phase in the system is (HD,MC) phase, which is specified by the following conditions:
\begin{equation} \label{HD,MC}
 \beta_{eff} < 1/2, \quad \beta_{eff} < \alpha ; \quad \beta > 1/2, \quad  \alpha_{eff} > 1/2.
\end{equation}
The particle currents and bulk densities in this phase are given by
\begin{equation}
J_{I}=\beta_{eff} (1-\beta_{eff}), \quad  \rho_{I,bulk}=1-\beta_{eff} , \quad J_{III}=1/4, \quad  \rho_{III,bulk}=1/2. 
\end{equation}
From Eq.(\ref{stat_current}) it can be easily shown that 
\begin{equation} 
\beta_{eff}=\frac{1}{2}-\frac{\sqrt{2}}{4}.
\end{equation}  
Comparing this result with the conditions (\ref{HD,MC}), we finally derive 
\begin{equation} 
\alpha > \frac{1}{2}-\frac{\sqrt{2}}{4}, \quad \mbox{and } \beta >\frac{1}{2}.
\end{equation}
These expressions describe the phase space for (HD,MC) phase.

Thus the analysis that considers TASEP on the lattices with junctions as coupled asymmetric exclusions systems suggests the existence of 5 phases. The calculated phase diagram is shown in Fig. 3. It can be seen that two phases, namely, (LD,MC) and (LD,HD), correspond to stationary phase boundaries. This result is due to the neglect of correlations around the junction point. However, at these conditions, the domain wall, that separates two phases can move with equal probability to the right or to the left \cite{kolomeisky98}. Then the density profiles will be different from the one predicted in the simplest theory, that treats the correlations around the junction in a mean-field manner. It can be expected that the density profile is linear in the chain I and the chain II segments for (LD,MC) phase, extrapolating between coexisting low-density and high-density phases, while the density profile in the chain III  still corresponds to the maximal-current phase. For (LD,HD) phase the linear density profiles can be found in all lattice segments, although the slopes  strongly depend on inter-segment density correlations.

\begin{figure}[h] 
\label{fig3}
\centering
\includegraphics[scale=0.5, clip=true]{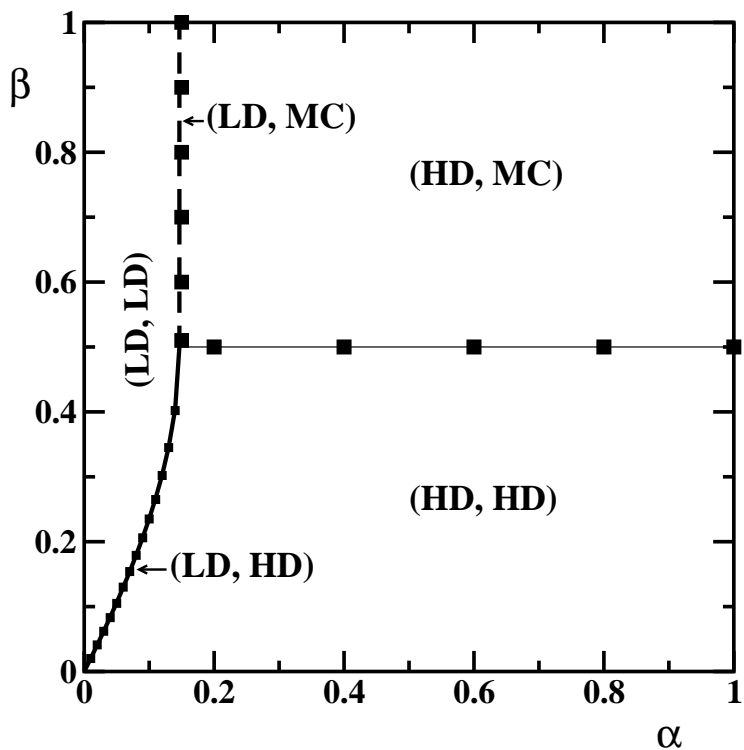}
\caption{\small{Phase diagram for the totally asymmetric exclusion system on the lattice with the junction. Lines are calculated from theoretical predictions for phase boundaries, while symbols are from Monte Carlo computer simulations. Thick solid line corresponds to a non-equilibrium first-order phase transition between (LD,LD) and (HD,HD) phases. Thin solid line shows a continuous phase transition between (HD,HD) and (HD,MC) phases. Meanwhile, thick dashed line describes a mixed phase transition between (LD,LD) and (HD,MC) phases: a first-order transformation in the chain I and II, and the continuous phase transition in the chain III.  The size of the symbols reflect the statistical error of computer simulations.}}
\end{figure}

\subsection{Phase boundaries}

Qualitative domain-wall arguments presented above indicate that particle density changes linearly for some  phase coexistence lines. However, exact  linear density profiles might differ significantly from the densities obtained within the simplest approximate theory that neglects the correlations near the junction.  Here we utilize the domain-wall approach \cite{ kolomeisky98} to account for these correlations in order to derive the phase boundary density profiles explicitly.

The line specified by $\alpha=\frac{1}{2}-\frac{\sqrt{2}}{4}$ for $\beta > 1/2$ (see Fig. 3) describes the mixed phase coexistence between (LD,LD) and (HD,MC). When crossing this line, the density jumps in the lattice segments I and II, while the change is continuous in the chain III. At the phase boundary the domain wall in the chain I or II moves randomly with equal forward and backward rates between the low-density ($\rho_{LD}=\frac{1}{2}-\frac{\sqrt{2}}{4}$) and high-density ($\rho_{HD}=1-\beta_{eff}=\frac{1}{2}+\frac{\sqrt{2}}{4}$) regions. The domain wall picture cannot be used in the lattice chain III when the maximal-current phase appears \cite{kolomeisky98}, and the inter-segment correlations are not significant. Then the  resulting density profiles in the left lattice segments are expected to be linear, connecting the low-density and high-density values.   

To determine the density profiles at the phase boundary between (LD,LD) and (HD,HD) phases is a much more complicated problem because of the strong correlations around the junction. This phase coexistence line is given by the following conditions:
\begin{equation}
\alpha< \frac{1}{2}-\frac{\sqrt{2}}{4}, \quad \beta  = \frac{1-\sqrt{1-8\alpha(1-\alpha)}}{2},
\end{equation}
as shown in Fig. 3. The domain wall  separates the coexisting stationary phases and it can be found in any lattice segment. Let us define a position of the domain wall in the system via relative coordinate $x$,
\begin{equation}
x=\frac{i}{L},
\end{equation}
where $i$ is the site index and $L$ is the length of one lattice segment. Therefore, the case of $0<x \le 1$ describes the chain I and II, while $1 < x \le 2$ corresponds to the chain III.

When the domain wall is in the chain I (and simultaneously in the chain II) it can move with the same rate $u_{I}$ to the left or to the right, while in the chain III it travels forward or backward with the rate $u_{III}$: see Fig. 4. These rates can be determined by utilizing the expression (\ref{dw_vel}),
\begin{equation}
u_{k}=\frac{J_{k}}{\rho_{+}^{k}-\rho_{-}^{k}}, \quad \mbox{ for } k=I, III,
\end{equation} 
where
\begin{equation} \label{rho+-}
\rho_{-}^{I}=\alpha, \quad \rho_{+}^{I}=1-\alpha,  \quad \rho_{-}^{III}=\beta, \quad  \rho_{+}^{III}=1-\beta,
\end{equation}
and
\begin{equation} 
J_{I}=\alpha (1-\alpha), \quad J_{III}=2J_{I}.
\end{equation}
As a result, we obtain the following expressions for the rates $u_{I}$ and $u_{III}$,
\begin{equation} \label{u1}
u_{I}=\frac{\alpha(1-\alpha)}{1-2\alpha},
\end{equation}
\begin{equation}\label{u3}
u_{III}=\frac{2\alpha(1-\alpha)}{1-2\beta}=\frac{2\alpha(1-\alpha)}{\sqrt{1-8\alpha(1-\alpha)}}.
\end{equation}

\begin{figure}[h] \label{fig4}
\centering
\includegraphics[scale=0.55, clip=true]{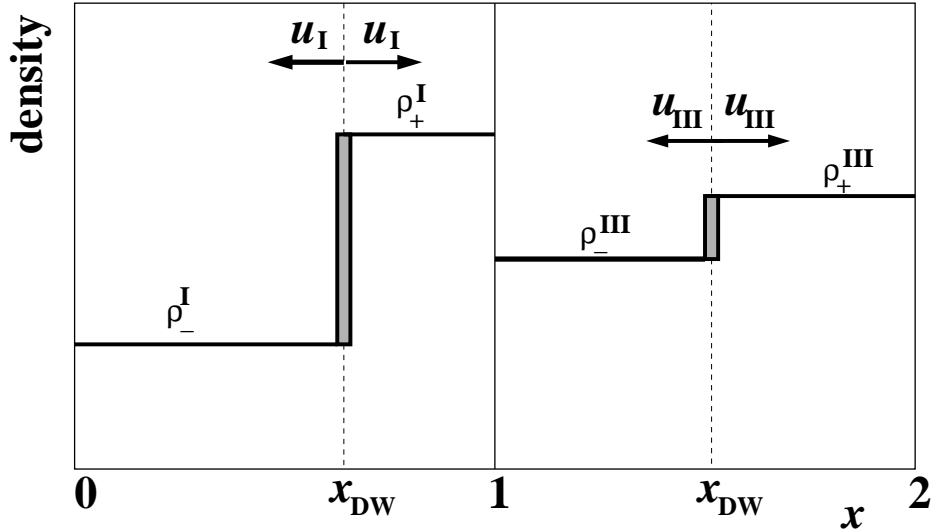}
\caption{\small{Schematic picture of the domain wall dynamics at the phase coexistence line between (LD,LD) and (HD,HD) phases. The domain wall in the chain I and II hops to the right or left with the rate $u_{I}$, while in the chain III it moves with the rate $u_{III}$.}}
\end{figure}

To calculate the density profiles we introduce a  probability $P_{I}$ to find the domain wall at any position in the chain I or II, and $P_{III}$ gives the probability that the domain wall is in the lattice chain III. These probabilities are obviously  normalized,
\begin{equation}\label{P_norm}
P_{I}+P_{III}=1.
\end{equation}
The probability that the domain wall occupies a specific site $i$ is equal to $P_{I}/L$  or $P_{III}/L$ for $i<L$ and $i>L$, respectively. Then at the junction 
\begin{equation}\label{P_prop}
u_{I}P_{I}/L=u_{III}P_{III}/L.
\end{equation}
This relation reflects the fact that the domain wall has equal probability to travel between different lattice segments. By combining last two equations we obtain:
\begin{equation}\label{P}
P_{I}=\frac{u_{III}}{u_{I}+u_{III}}, \quad   P_{III}=\frac{u_{I}}{u_{I}+u_{III}}.
\end{equation}
These expressions have a simple physical explanation. The domain wall spends less time in the lattice segments where it fluctuates faster. All lattice chains have the same length $L$, and the domain wall in the lattice segment where it fluctuates faster is able to diffuse to the junction point quicker. As a result, it will jump to another lattice chain more frequently. This is a reason for strong  inter-segment density correlations at the first-order phase transitions.

We can determine the probabilities of having the domain wall at any  position less then a certain value of x. If the domain wall is in the channels I and II at the coordinate $x_{DW}$, then this probability  is given by
\begin{equation} \label{prob_dw1}
Prob(x_{DW}<x)=P_{I} x, \quad 0<x \le 1,
\end{equation}
Similarly, for the domain wall in the lattice chain III,
\begin{equation} \label{prob_dw2}
Prob(x_{DW}<x)=P_{I}+P_{III} (x-1), 1<x \le 2.
\end{equation}
Then the  density at any position  can be calculated as
\begin{equation}
\rho(x) = \rho_{-}^{k} Prob(x_{DW} > x)+\rho_{+}^{k} Prob(x_{DW} < x), \quad k=I, III.
\end{equation}
Finally, combining  Eqs. (\ref{rho+-}), (\ref{u1}), (\ref{u3}), (\ref{P}), (\ref{prob_dw1}) and  (\ref{prob_dw2}) we obtain
\begin{equation}\label{lin_dens1}
\rho(x)_{I} = \alpha+\frac{2(1-2\alpha)^{2}}{2(1-2\alpha)+\sqrt{1-8\alpha(1-\alpha)}} x,  \quad  0 < x \le 1;
\end{equation}
and
\begin{eqnarray}\label{lin_dens2}
\rho(x)_{III} & = &\frac{1-\sqrt{1-8\alpha(1-\alpha)}}{2}+\frac{2(1-2\alpha)\sqrt{1-8\alpha(1-\alpha)}}{2(1-2\alpha)+\sqrt{1-8\alpha(1-\alpha)}} \nonumber \\ 
  &  & + \frac{1-8 \alpha(1-\alpha)}{2(1-2\alpha)+\sqrt{1-8\alpha(1-\alpha)}} (x-1), 
\end{eqnarray}
for $1 < x \le 2$. At the entrance sites  we have, as expected, $\rho(x=0)_{I} =  \alpha$, while at the last site $\rho(x=2)_{III} = \frac{1+\sqrt{1-8\alpha(1-\alpha)}}{2} =1-\beta$.  At the junction, the densities are equal to
\begin{eqnarray}\label{lin_dens}
\rho(x=1)_{I}   &  =  & \alpha+\frac{2(1-2\alpha)^{2}}{2(1-2\alpha)+\sqrt{1-8\alpha(1-\alpha)}}, \nonumber \\
\rho(x=1)_{III} &  =  & \frac{1-\sqrt{1-8\alpha(1-\alpha)}}{2}+\frac{2(1-2\alpha)\sqrt{1-8\alpha(1-\alpha)}}{2(1-2\alpha)+\sqrt{1-8\alpha(1-\alpha)}}. 
\end{eqnarray}
It is important to note  that the densities at junction are not equal to the values $1-\alpha$ for the chain I and II, and $\beta = \frac{1-\sqrt{1-8\alpha(1-\alpha)}}{2}$ for the chain III, respectively, as expected from the simplest approximate theory that  assumes  an independent coupling between the lattice segments. The domain-wall approach allows  to take into consideration the correlations in the densities around the junction.

\section{Monte-Carlo simulations and discussions}

In order to check the validity of our approximate theory we performed  extensive computer Monte Carlo simulations. Since the computer calculations for the cases of phase transitions are very time-consuming, especially for small values of entrance and exit rates, we utilized one of the continuous-time Monte Carlo algorithms, the so-called  BKL algorithm,  first introduced  by Bortz, Kalos and Lebowitz almost 30 years ago \cite{BKL}. The main idea of the BKL algorithm is to create an event-driven update scheme, so that the rejected, "eventless" steps are skipped. The method is simple to implement, and with the latest improvements \cite{BKL2} it is also very efficient and fast.

In our simulations the number of effective steps per site  was typically around $10^{7}$. At phase transitions we ran simulations  much longer, and the number of effective steps per site was $10^{8}$-$10^{9}$. For all simulations we neglect first $3\%$ of Monte Carlo steps to account for the time that the system takes to achieve a stationary state. Our theoretical calculations  assume infinite size lattice segments, however, in our simulations we used $L=100$ and we checked that for larger sizes of lattice segments the results do not deviate from the ones presented here. 

A phase diagram obtained from Monte Carlo simulations is shown in Fig. 3. The boundaries between the stationary phases have been determined by considering the saturation in the currents and comparing qualitative changes in the  density profiles. Specifically, the phase coexistence line between (LD,LD) and (HD,HD) phases is found when the density profiles become linear. The boundary between (LD,LD) and (HD,MC) phase is determined when the linear density profile is observed in the chains I and II, and the overall particle current saturates. Similarly, the current saturation method allows to specify the phase boundary between (HD,HD) and (HD,MC) phases. The overall error in the determination of phase boundaries is less than $5\%$ \cite{pronina04}. To illustrate our approach, the dependence of current on the entrance rate $\alpha$ (for fixed exit rate $\beta=1$) is plotted in Fig. 5. It can be clearly seen that the current  saturates at $\alpha \approx 0.15$.

\begin{figure}[h] \label{fig5}
\centering
\includegraphics[scale=0.3, clip=true]{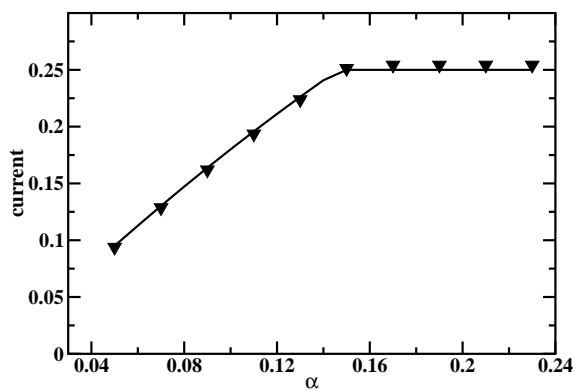}
\caption{\small{The particle current as a function of the entrance rate $\alpha$. The exit rate is fixed, $\beta=1$. The current saturates at $\alpha \approx 0.15$ that corresponds to the phase transition between (LD, LD) and (HD, MC) phases.  A solid line describes  theoretical predictions, while symbols are the  results of computer simulations.}}
\end{figure}

Monte Carlo simulations allowed  also to calculate explicitly the particle densities. The resulting density profiles for different stationary phases are shown in Fig. 6. Since the stationary properties  in the chains I and II are essentially the same, we  investigated in detail only one of two equivalent lattice segments. The computer simulations for the densities in the bulk phases are well described by the theoretical predictions. Note, however, the deviations near the junction that grow as the system  approaches the phase boundaries.
\begin{figure}[tbp]
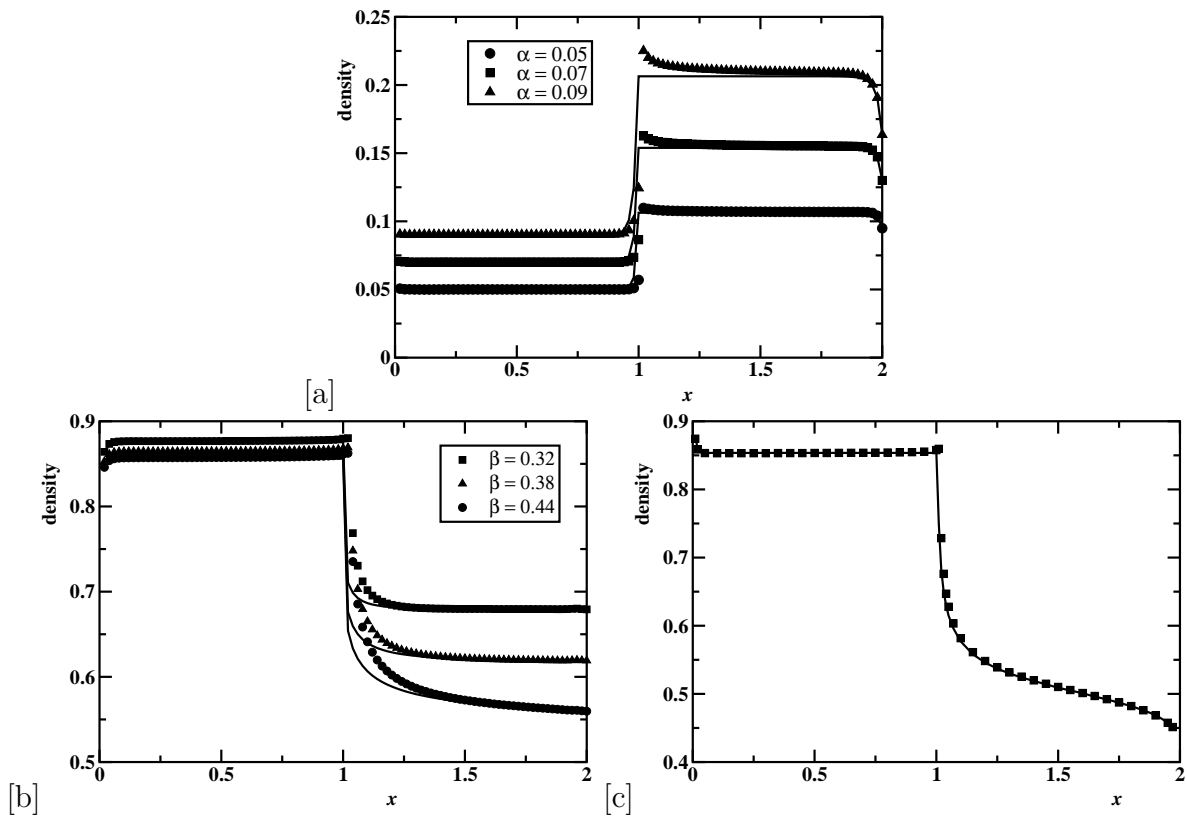
\label{fig6}
\centering
[a]\includegraphics[scale=0.3,clip=true]{Fig6a.junction.eps}
[b]\includegraphics[scale=0.3,clip=true]{Fig6b.junction.eps}
[c]\includegraphics[scale=0.3,clip=true]{Fig6c.junction.eps}
\caption{Density profiles for different bulk stationary phases: (a) (LD,LD) phase with $\beta=1$ and three different entrance rates $\alpha=0.05$, 0.07 and 0.09; (b) (HD,HD) phase with $\alpha=0.8$ and three different exit rates $\beta=0.32$, 0.38 and 0.44; (c) (HD,MC) phase with $\alpha=1$ and $\beta=0.58$. Lines are our theoretical predictions for the full-length density profiles calculated from $\alpha$, $\beta_{eff}$, $\alpha_{eff}$ and $\beta$ by using the exact expressions derived in Ref. (\cite{derrida93}). Symbols are obtained from the Monte Carlo computer simulations.}
\end{figure}

 The situation is very different for the phase coexistence line between (LD,LD) and (HD,HD) phase. Our theoretical approach predicts linear density profiles in all lattice segments, although with different slopes. The extensive computer simulations (see Fig. 7) mainly confirm these suggestions, although the density profiles in the chains I and II deviate slightly from the linearity. There are possible sources of these deviations: 1)  it might be  due to the errors in the determination of the exact position of this phase boundary; and/or 2) the cross-correlation between the particles in the chains I and II as was observed for the related system \cite{brankov04}.

\begin{figure}[ht]
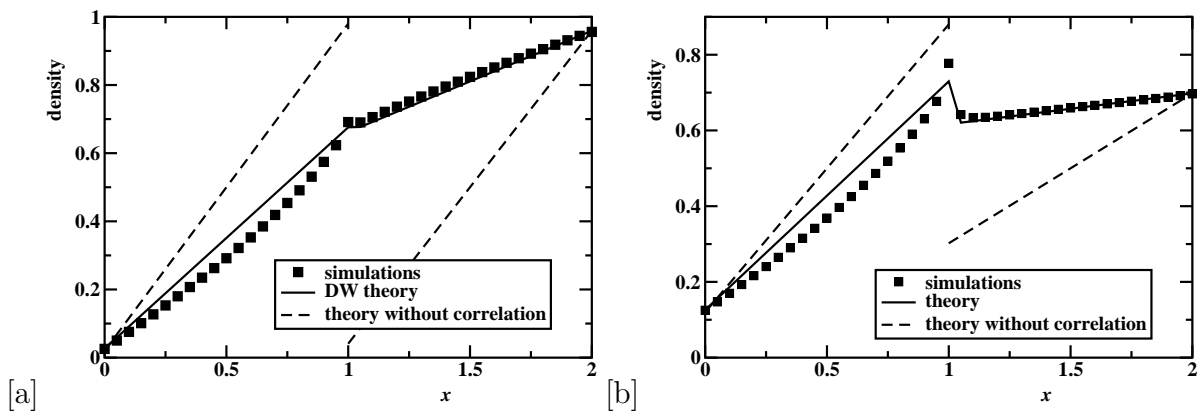
 \label{fig7}
[a] \includegraphics[scale=0.3, clip=true]{Fig7a.junction.eps}
[b] \includegraphics[scale=0.3, clip=true]{Fig7b.junction.eps}
\caption{ \small{Density profiles  for phase coexistence line between (LD,LD) and (HD,HD) phases for different parameters: (a) $\alpha = 0.02$, $\beta=0.0408$; (b)  $\alpha = 0.12$, $\beta=0.3020$. Solid lines are theoretical predictions from the modified domain-wall theory, namely,  Eqs. (\ref{lin_dens1}) and (\ref{lin_dens2}). Dashed lines correspond to the predictions of the simplest theory without correlations.}}
\end{figure} 

The analysis of stationary properties of TASEP on the lattice with junction indicates an excellent  agreement with the theoretical predictions: see Figs. 3, 5, 6 and 7. Although the simplest approximate theory neglects the correlations around the junction, it does not strongly affect the position of phase boundaries, stationary currents and bulk density profiles. However, the effect of correlations is important at the first-order phase transitions between (LD,LD) and (HD,HD) phases. The extension of the domain-wall phenomenological approach, that argues that the domain wall fluctuates with different rates in the different lattice segments, is able to account for density correlations as compared with the results from the computer simulations (Fig. 7).

\section{Summary and conclusions} 

The stationary properties of totally asymmetric simple exclusion processes on the lattices with junctions are investigated with the help of the simple approximate theory and by analyzing extensive computer Monte Carlo simulations. It is found that the phase diagram of the system consists of three stationary phases. This behavior is similar to asymmetric exclusion processes on the lattices without junctions, although the maximal-current phase cannot be sustained in the lattice branches before the junction due to the stationary current limitations. 

There are three different types of phase transitions in this system. On the coexistence line between (LD,LD) and (HD,HD) phase there are density jumps in all lattice segments that correspond to the first-order phase transitions. On the phase boundary between (HD,HD) and (HD,MC) the density profile changes continually in the lattice chain III. However, the phase transition between (LD,LD) and (HD,MC) has quite an unusual mixed character: the first-order transformation in the lattice chain I and II, and the continuous change in the lattice chain III. This phase behavior differs significantly from TASEP on the lattices without junctions.

To analyze the dynamics of asymmetric exclusion processes on the lattice with the junction the approximate theoretical approach has been developed. According to this theory the system of particles moving on the lattice with the junction can be viewed as three TASEP on the lattices without defects that coupled at the junction. As the simplest approximation, the correlations near the junction position are neglected. This  method allows to calculate explicitly all stationary properties and the phase diagram. It is found that the theoretical predictions are in excellent agreement with Monte Carlo computer simulations for all phase regions except for the phase coexistence line between (LD,LD) and (HD,HD) phase. 

For the first-order phase boundary, where the inter-segment density correlations are important,  the modified domain-wall  approach is developed. We argue that the domain wall, that separates the low-density and high-density phases in each lattice segment, fluctuates with the different rates in the different lattice branches. It means that the domain wall does not spend the same time in all lattice segments, and this leads to the correlations observed in the system. The computer Monte Carlo simulations fully support  the predictions from the modified domain-wall theory. It is suggested that this approach is general enough to be used successfully to account for correlations  in other inhomogeneous asymmetric exclusion processes. For example, for  TASEP on the lattices with  double-chain sections \cite{brankov04}, i.e., the system with two junctions, can explicitly predict the density profiles and it can explain the observed density correlations.          

There are several extensions of the system that can be explored in a future. In the original model, the particle dynamics in the lattice segments before the junction have been identical. It will be interesting to investigate a system where two lattice branches before the junction are dynamically different. More complex dynamics is expected for a system where more than three lattice branches are joined together at the junction point. It is expected that our approximate theory without correlations and the modified domain-wall approach, supported by computer Monte Carlo simulations, will provide a reasonable way of analyzing these complex non-equilibrium systems.

\ack

The support from the Camille and Henry Dreyfus New Faculty Awards Program (under Grant No. NF-00-056), from the Welch Foundation (under Grant No. C-1559), and from the US National Science Foundation through the grant CHE-0237105 is gratefully acknowledged.

\end{document}